\documentstyle[epsfig,float,twocolumn,prb,aps]{revtex}

\renewcommand{\S}{{\vec S}}

\begin{document}
\draft

 \twocolumn[\hsize\textwidth\columnwidth\hsize\csname @twocolumnfalse\endcsname

\title{The Haldane gap for the $S=2$  antiferromagnetic Heisenberg   chain
revisited}

\author{
Xiaoqun Wang$^{1,2}$, Shaojing Qin$^3$ and Lu Yu$^{4,3}$
}
\address{
$^1$Institut Romand de Recherche Numerique en Physique Des
Materiaux (IRRMA), PPH-333, EPFL, CH-1015 Lausanne, Switzerland
 }
\address{
$^2$ Max-Planck-Institut f\"ur Physik Komplexer Systeme,
N\"othnitzer Str. 38, D-01187 Dresden
}
\address{
$^3$Institute of Theoretical Physics, P. O. Box 2735, Beijing
100080, P.R. China
}
\address{
$^4$International Center for Theoretical Physics, P. O. Box 586,
34100 Trieste, Italy
}

\date{\today}  
\maketitle
\begin{abstract}
Using the density matrix renormalization group (DMRG) technique, we carry out a
large scale numerical
calculation for the $S=2$ antiferromagnetic Heisenberg chain. Performing
systematic  scaling analysis
for both the chain length $L$ and the number of optimal states kept in the
iterations  $m$, the Haldane gap  $\Delta (2)$ is estimated accurately as $(0.0876\pm0.0013)J$.
Our systematic analysis for the $S=2$ chains not only ends the controversies 
arising from various DMRG calculations and Monte Carlo simulations, but
also sheds light on how to obtain reliable results from the DMRG calculations
for other complicated systems. 
\end{abstract}

 ]

Since Haldane\cite{Haldane 1983} conjectured that integer-spin $S$
antiferromagnetic Heisenberg (AFH)
chains have a gap
$\Delta(S)$ in the excitation spectrum, much work \cite{Affleck
1987,Affleck 1989,Kennedy,%
Sorensen,White 1993,Golinelli}
 has been done on  spin one 
Heisenberg AFH chains. In particular, the valence bond solid  (VBS) model has
been proposed to
provide physical interpretation for  this important property.\cite{Affleck 1987} In recent
years, very accurate
 estimates of the gap [$\Delta(1)=0.41049(2)J$]
have been obtained by White and Huse,\cite{White 1993}  and by  Golinelli
{\it et al},\cite{Golinelli}
using the density matrix
renormalization group (DMRG) method\cite{White 1992} and exact diagonalization
 with proper extrapolation, respectively.
Experimental evidence for the gap has also been clearly shown on some
quasi-one-dimensional $S=1$ AFH
materials, such as Ni(C$_2$H$_8$N$_2$)$_2$NO$_2$(ClO$_4$)
(NENP),\cite{Gran}
Y$_2$BaNiO$_5$,\cite{no,Lu}
and AgVP$_2$S$_6$.\cite{AgV}
Moreover, nontrivial degenerate ground states
show up when the chain becomes
open. This property is related to  the impurity effects  observed in
Y$_{2-x}$Ca$_{x}$BaNiO$_5$.\cite{Dit}
As pointed out by one of us, the boundary effects and associated impurity
effects distinguish the low-energy properties of the Haldane system from
other quasi-one-dimensional  gapped spin systems.\cite{Wang 1998}

The $S=2$ AFH chain, beyond its own interest, is essential for  further
verification of the Haldane
conjecture. In the limit of large $S$ the spin chains become
quasi-classical, which
implies that the difference between the half-integer and integer spin
chains, {\it e.g.}
 the absence/presence of the Haldane gap,
must  diminish. The  $S=2$ Heisenberg AFH chain is thus the first case to
check this behavior. Recently there appears to be also experimental
evidence for a Haldane gap
in an $S=2$ system.\cite{Granrot} The material under study  is
(2,2'-bipyridine)trichloromanganese(III),
MnCl$_3$(bipy). The manganese ions form effective $S=2$ spins and are
coupled to a quasilinear
chain by chlorine ions. The Haldane gap has been measured by
determining    the low
temperature behavior, and it was estimated as  $0.07J$.

For the  $S=2$ antiferromagnetic Heisenberg chain described by
\begin{equation}
H=J\sum_{i=1}^L\S_{i}\cdot\S_{i+1},
\end{equation}
 numerical calculations are much more elaborate than for  the $S=1$ case.
There are two reasons. The first one is that the value of the Haldane gap
for $S=2$ is  roughly one order
of magnitude smaller than that of $S=1$ chain. Correspondingly, the
correlation length in the ground
state also increases roughly by one order of magnitude.
Consequently much longer chains are required to reach the convergent regime
with respect
to the chain length. In fact, to  exclude  the finite size effects one has
to go up to chain lengths of  about one thousand for open boundary conditions.  
The second reason is that the number of degrees of freedom per spin is
five for $S=2$ instead of three for $S=1$, so the exact diagonalization can
be carried only for rather
short chains (of the order 10, much shorter than the correlation length).

Nevertheless, the $S=2$ Heisenberg AFH chain has been numerically studied by
quite a few groups using
different Monte Carlo techniques and DMRG methods,\cite{Deisz,Hatano,Qin
1995,Sun,Nishiyama,%
Yamamoto,Schollw 1995,Yamamoto 1996,Qin 1997,Qin 1998,Weis,Troyer,Schollw
1998}   and  we have  listed
 results  of these calculations chronologically in Table I.  From these
values, on the other hand,
it is not easy to extract a reliable
estimate of the Haldane gap for the $S=2$ chain. Recently, Kashurnikov {\it
et al},\cite{Troyer}
based on their ``Worm'' Monte Carlo (WMC) results (which are excellent for
$S= 1/2 ,1 $ cases)
expressed some doubt on the value of  $\Delta(2)$  obtained by  DMRG
calculations.
This has indeed independently raised a serious question concerning both the capability
of the DMRG method itself and
the true value of the Haldane gap for the $S=2$ case.
It seems true that  analysis based on  a single DMRG calculation with fixed
number of
kept states $m$ does not lead  to  a conclusive answer.
In this paper, we carry out a large scale DMRG calculation with up to one
thousand and two hundred sites as well as a systematic
scaling analysis  for  the number of the kept   optimal states, up to 400.
We will show then that $\Delta(2)=0.0876J\pm 0.0013J$.
In addition, this analysis  also sheds some light on how   to judge
in which case a DMRG calculation and the corresponding scaling analysis can
be trusted.

{
\table
\caption{Values of the Haldane gap $\Delta$ for $S=2$ Heisenberg AFH chain
as calculated by various groups using different techniques.
WLMC: World line Monte Carlo method; PMC: Projection Monte Carlo Method;
LCMC: Loop cluster Monte Carlo method; WMC: Worm Monte Carlo method.
BCs: boundary conditions; PB (OB): periodic (open) boundary conditions.
$m$: the number of states kept in the DMRG iterations. SA indicates
whether or not a scaling analysis has been  used to determine the gap
value. $L$ indicates the largest size of the chains in each case.
The numbers in parentheses are errors, for the gap at last digit(s),
given in those references. 
In the following table, we set $J=1$. \label{table1}}
\begin{tabular}{lllccc}
methods    & $\Delta$         & $L$        & BCs& SA & $m$\\
\hline
WLMC\cite{Deisz}       & $0.08$          & $64$       & PB  & No.   &--\\
DMRG\cite{Hatano}       & $0.02$          & $100$  & OB  & Yes.  &120\\
DMRG\cite{Qin 1995}       & $0.055(15)$     & $70$    & OB  & Yes.  &110\\
PMC\cite{Sun}        & $0.05$          & $32$       & PB  & No.   &--\\
DMRG\cite{Nishiyama}     & $0.055(15)$  & $40$    & OB  & Yes. & 110\\
WLMC\cite{Yamamoto}       & $0.049$         & $ 128$ & PB  & Yes.  &--\\
DMRG\cite{Schollw 1995}       & $0.085(2)$      & $ 350$    & OB  & Yes.
&300\\
WLMC\cite{Yamamoto 1996}  & $0.074(16) $    & $512$   &PB   &Yes. &--\\
DMRG\cite{Qin 1997}       & $0.082(3)$      & $150$   & OB  & Yes.  &250\\
DMRG\cite{Qin 1998}       & $0.085(1)$      & $80  $     & PB  & Yes.  &1700\\
LCMC\cite{Weis}       & $0.09(1) $      & $400$     & PB  & No.   &--\\
WMC\cite{Troyer}        & $0.1032(7)$     & $500 $     & PB  & No.   &--\\
DMRG\cite{Schollw 1998}       & $0.0907(2)$      & $600$      & OB  & Yes.
&400
\end{tabular}
}

\vspace{3mm}
Let us first recall   some related physical properties. First of all,
as emphasized by Schollw\"ock {\it et al}\cite{Schollw 1995} and Qin {\it
et al},\cite{Qin 1995}
for the  periodic boundary conditions, the ground state is a singlet,
whereas for open boundary conditions
the topological edge excitations  make $S^{tot}=0,1,....$  states
non-degenerate at finite lengths. Of course,
in the thermodynamic limit these states become degenerate,
and  for relatively short chains, the interaction between these
topological excitations has to be carefully taken
into account in the scaling analysis.\cite{Qin 1995} Another way of removing
the ``surface'' effects
is to introduce different spin values and coupling constants at the
edges.\cite{Schollw 1995}
In our numerical calculations following
Schollw\"ock {\it et al},\cite{Schollw 1995} we attached two $S=1$ spins
to screen the edge
spins at both ends of an open $S=2$ chain.
This trick is  helpful in practice for an accurate evaluation of the
Haldane gap, although
the spin gap as a bulk quantity should be independent of the boundary
conditions.
The main reason is that the surface energies
with different $S^{tot}_z$ are  affected by truncations differently than
the bulk properties.

As follows from the one dimensional field theory,
the leading finite size correction to the Haldane gap with  open
boundary conditions
is proportional to the inverse of the length squared\cite{Sorensen}
\begin{equation}
\Delta(m=\infty, L)=\Delta(S)+\frac {v^2\pi^2}{2\Delta(S)
L^2}+{\cal O}(\frac 1 {L^3})
\end{equation}
where $v$ is the spin wave velocity, and $\Delta(S)$ the value of
the Haldane gap
in the thermodynamic limit.  This formula shows that the true gap
should be  a minimum of the parabola as a function of the inverse
chain length $L^{-1}$, if all states are kept ($m= \infty$) and higher
order terms are negligible.
In the DMRG calculations, however, the truncation of the Hilbert space leads to
 deviations from the above  asymptotic behavior. The deviation is not so
serious for the spin  1/2  or spin 1 systems, which are usually not
explicitly emphasized. For more complicated systems, like the $S=2$ case,
the deviation is substantial,
especially when $m$ is not sufficiently  large.

\begin{figure}[ht]
\epsfxsize=3.3 in\centerline{\epsffile{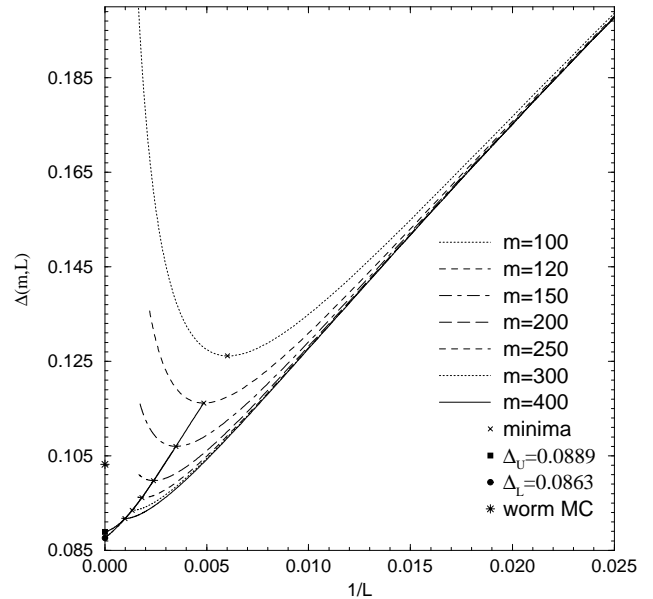}}
\vspace{0.5cm}
\caption[]{The gap $\Delta(m,L)$ as a function of the number of states $m$
kept in the DMRG
calculations and the chain length $L$. Each cross for a given $m$ indicates
the position
of the minimum of $\Delta(m,L)$. The solid square denotes the  upper
estimated value $\Delta_U$, while the
solid circle denotes the lower estimated value $\Delta_L$. The star on the
vertical axis indicates the
recent  WMC result.\cite{Troyer}}
\label{Fig}
\end{figure}

In Fig. 1, we show the gap $\Delta(m,L)$,
 as measured by  the difference between the lowest
energy states with $S_z^{tot}=1$ and  $S_z^{tot}=0$, respectively,
as a function of the chain length $L$ and the number of states $m$ kept in
the DMRG iterations. One can immediately see from Fig. 1 that
there exists a minimum as denoted by the cross for any given number of kept
states $m$.
Therefore, in principle  one cannot use the extrapolation with respect to
$1/L$ in order to get the gap value for the thermodynamic limit when $m$
is not sufficiently large. As pointed out by Schollw\"ock {\it et al},
when the chain length is not long enough, the scaling is essentially
dominated by the $1/L$ term . This issue has also been  addressed by Qin
{\it et al}.\cite{Qin 1997}
Taking this into account one may still obtain  a reasonable
estimate of the gap value  for the open chains by  
keeping a sufficiently large number of
states, as  done
by Schollw\"ock {\it et al},\cite{Schollw 1995} Qin  {\it et al}\cite{Qin
1997} and
Schollw\"ock.\cite{Schollw 1998} Their obtained values are comparable with
our present
results shown in Fig.1.

However,  in order to obtain a
reliable value for the thermodynamic limit, one has  to make some further
analysis as illustrated in Fig. 1. It is obvious that the minimum of each
curve for $\Delta(m,L)$
shifts towards the vertical axis, as $m$ increases. In the limit $m
\rightarrow \infty$
the minimum should be right on the vertical axis, as follows from Eq.(2).
 In fact, the convergence of
minimum location  with respect to $m$ is quite fast.
The minimum value of $\Delta(m=400,L)$ is $0.0917385344$ at $L=1010$.\cite{digits}
We have fitted these positions for $m=120,150,200,250,300,400$
with respect to $1/L$. We note that the extrapolation is performed for 
$m$ and $L$ simultaneously.\cite{extra}
The values $\Delta_L$ and $\Delta_U$ correspond to two different extrapolations
using the polynomial fit and the spline fit, respectively.
We have also checked that  the results change very little, by leaving out
the minimum
for $m=120$ or by including more minima  in the range $150\leq m\leq 400$.
These changes are
small compared with the difference between $\Delta_U$ and $\Delta_L$.

Therefore, we can safely conclude that  the Haldane gap is
$0.0876\pm0.0013$ for the $S=2$ AFH
chain, where the uncertainty comes from the simultaneous extrapolation only.
Our analysis is based on the comparison of the one dimensional 
field theory asymptotic and our actual DMRG results.

It is  interesting to see how far are the values of the Haldane gap
obtained by the DMRG
method
for $S=1$ and $S=2$ AFH chains from those given by the following asymptotic
formula for large $S$ 
in the non-linear $\sigma$ model:\cite{Haldane 1983}
\begin{equation}
\Delta(S)=\alpha S(S+1){\rm e}^{-\pi\sqrt{S(S+1)}},~~~{\rm for ~} S\gg1
\end{equation}
If we fix the numerical  factor $\alpha$ using the value for
$S=1$, then we obtain $\Delta(S=2)=0.0470$, which is only $55\%$ of
our value. We therefore conclude that the above equation is only  a
semiquantitative,  asymptotic formula.

Before concluding we would like to make two remarks regarding
the  numerical calculations:

i) It is well-known that Monte Carlo techniques involve both statistical
errors and
systematic errors. In fact, the systematic errors must be
{\it efficiently} eliminated before analyzing the statistical errors.
For an {\it accurate} evaluation of quantities such as the ground state
energy and energy
gap, it is usually difficult to ensure to which  extent the
configuration dependence is sufficiently eliminated in reducing the
systematic errors.
It is very nice that the Worm Monte Carlo simulations have produced
excellent results for $S=1/2, 1$
 AFH chains, reproducing even the ``relativistic'' dispersion
relations\cite{Qin 1998} for $S=1$ case.
 The same method was used to obtain  $\Delta(2)=0.1032(7)$, which is quite
different from our
result.  We would like to mention that the excellent results in $S= 1/2, \
1 $ cases
cannot guarantee that the systematic errors have been efficiently excluded for
the $S=2$ case, since the configuration dependency involved in these cases
is quite different. This relevant issue for the Monte Carlo simulations
was also addressed in Ref. \onlinecite{Schollw 1996}.

ii) The DMRG calculations  involve a systematic error, too. It
originates from
the truncation of the  renormalized Hilbert space. In relatively  simple
applications, like
$S=\frac 1 2$ and $S=1$ Heisenberg chains, these errors are not   serious,
while for  more
complex systems, such as $S=\frac 3 2$ and $S=2$ chains, they  become
rather crucial.
With more and more applications of  DMRG to  complicated systems, we note that
the analysis based on DMRG calculations becomes reliable for the
thermodynamic limit
\begin{itemize}
\item[1)] when a given  set of numerical data follows very well the analytic
scaling up to a convergent regime
of scaling variable (usually $1/L$);
\item[2)] when various sets of numerical data
are computed for
a number  of $m$ like what we have done for the $S=2$ chain, and a further
scaling for $m$
is being carefully carried out. 
\end{itemize}
For the first case, we quote two more
examples. One is for a $S=3/2$  periodic chain with m=1200 sates
kept.\cite{karen32,wang1}
The analytical scaling behavior is available from the conformal filed theory,
and a careful scaling analysis  has shown that $S=3/2$ spin belongs to the
same universality class, as $S=1/2$ case  with  central charge $c=1$ and the
scaling exponent $\eta =1$.
 Another example is   an $S=2$ periodic chain,\cite{Qin 1998}  and the
number of states
was kept up to $m=1700$.\cite{wang1} Again, it has been shown unambiguously that $S=2$
chain
belongs to the same universality class as $S=1$ chain. In  both cases, the
results for the low-lying energy
excitations and  the correlation functions fit the scaling functions very
nicely in the convergent regime
of the scaling variables.  We can conclude that these DMRG results, like
the present one for the
$S=2$ Haldane gap,  are reliable.

Xiaoqun Wang is grateful to Steffen Mallwitz for the earlier collaboration, 
Ulrich Schollw\"{o}ck for helpful communication and Reinhard Noack for conversation.
Shaojing Qin is partially supported by NSFC and CAS.


\begin{references}
\bibitem{extra} Usually, one would expect to make two individual extrapolations:
one is made with respect to $m$ for each given $L$ and then the other is for $L$. 
This way is not realistic as seen from Fig. 1 and can involve more uncertainty of 
the extrapolations.
\end{references}
\end{document}